\documentclass[draftcls,onecolumn,romanappendices]{IEEEtran}

\usepackage[latin1]{inputenc}
\usepackage[english]{babel}
\usepackage{cite, graphicx}
\usepackage{amsmath, amsfonts, amssymb, float, latexsym, stackrel}
\usepackage{epstopdf}

\usepackage{notation}

\newcommand{\refE}[1]   {(\ref{eqn:#1})}
\newcommand{\refS}[1]   {Section~\ref{sec:#1}}
\newcommand{\refF}[1]   {Fig.~\ref{fig:#1}}
\newcommand{\refFig}[1] {Figure~\ref{fig:#1}}

\begin{document}

\title{Multi-Class Source-Channel Coding}

\author{Irina~E.~Bocharova, ~\IEEEmembership{Senior Member,~IEEE,}
        Albert~Guill{\'e}n~i~F{\`a}bregas,~\IEEEmembership{Senior Member,~IEEE,}
        Boris~D.~Kudryashov, ~\IEEEmembership{Senior Member,~IEEE,}
        Alfonso~Martinez,~\IEEEmembership{Senior Member,~IEEE,}
        Adri\`a~Tauste~Campo,~\IEEEmembership{Member,~IEEE,}
        Gonzalo~Vazquez-Vilar,~\IEEEmembership{Member,~IEEE}%
\thanks{I. E. Bocharova and B. D. Kudryashov are with the St. Petersburg University of Information Technologies, Mechanics and Optics, St. Petersburg 197101, Russia (e-mails: \{irina, boris\}@eit.lth.se).
A. Guill\'en i F\`abregas, A. Martinez and A. Tauste Campo are with the Department of Information and Communication Technologies, Universitat Pompeu Fabra, Barcelona 08018, Spain (e-mails: \{guillen, alfonso.martinez\}@ieee.org; adria.tauste@upf.edu). A. Guill\'en i F\`abregas is also with the Instituci\'o Catalana de Recerca i Estudis Avan\c{c}ats (ICREA), Barcelona 08010, Spain, and the Department of Engineering, University of Cambridge, Cambridge  CB2 1PZ, U.K. A. Tauste Campo is also with the Hospital del Mar Medical Research Institute, Barcelona 08003, Spain.
G. Vazquez-Vilar was with the Department of Information and Communication Technologies, Universitat Pompeu Fabra, Barcelona 08018, Spain. He is now with the Signal Theory and Communications Department, Universidad Carlos III de Madrid, Legan\'es 28911, Spain, and with the Gregorio Mara\~n\'on Health Research Institute, Madrid 28007, Spain (e-mail: gvazquez@ieee.org).}
\thanks{This work has been funded in part by the European Research Council under ERC grant agreement 259663, by the European Union's 7th Framework Programme under grant agreements 303633 and 329837 and by the Spanish Ministry of Economy and Competitiveness under grants RYC-2011-08150, TEC2012-38800-C03-03, TEC2013-41718-R and FPDI-2013-18602.
This work was presented in part at the 2014 IEEE Symposium on Information Theory, Honolulu, HI, June 29--July 4, 2014, and at the 8th International Symposium on Turbo Codes and Iterative Information Processing, Bremen, Germany, Aug. 18--22, 2014.}
}

\maketitle
\begin{abstract}
This paper studies an almost-lossless source-channel coding scheme in which source messages are assigned to different classes and encoded with a channel code that depends on the class index. The code performance is analyzed by means of random-coding error exponents and validated by simulation of a low-complexity implementation using existing source and channel codes. While each class code can be seen as a concatenation of a source code and a channel code, the overall performance improves on that of separate source-channel coding and approaches that of joint source-channel coding when the number of classes increases.
\end{abstract}


\section{Introduction}\label{sec:intro}

Reliable transmission of a source through a communication channel can be achieved by using separate source and channel codes, as shown by Shannon's source-channel coding theorem~\cite{Shannon48}. This means that a concatenation of a (channel-independent) source code followed by a (source-independent) channel code achieves vanishing error probability as the block length goes to infinity, as long as the source entropy is smaller than the channel capacity~\cite{Shannon48}.
However, in the non-asymptotic regime joint source-channel codes can perform strictly better.
This improvement (i.e.~reduction in error probability) has been quantified in terms of error exponents~\cite{Gall68,Csis80} and in terms of source and channel dispersion~\cite{wik11,kostina13}. Joint design has an error exponent at most twice of that of separate codes~\cite{Zhong06}, and a dispersion gain that depends on the target error probability; for vanishing values of the latter, the dispersion of joint design is at best half of the dispersion of separate design~\cite{kostina13}. This potential gain justifies the interest in practical finite-length joint source-channel codes.

Several practical joint source-channel coding schemes have been considered in the past. One possible approach is to adapt existing channel coding techniques to exploit the knowledge on the source statistics at the decoder side. Examples include a modification of the  Viterbi decoding algorithm to use the \textit{a priori} probabilities of the source bits \cite{Hagenauer95}, punctured turbo-codes with a modified iterative decoder \cite{GZ2002}, and source and channel LDPC codes with a decoder exploiting the joint graph structure of the codes and the source~\cite{FPPV2010}. Other schemes exploit the source statistics both at the encoder and decoder. In~\cite{SBAW2006}, source bits are matched to a non-systematic LDPC code via scrambling or splitting. In \cite{BH2000,HB2001,guyader2001} the authors propose a trellis-structure description of the Huffman code and an appropriate channel code so that joint decoding is possible. This technique has been extended  to arithmetic~\cite{grangetto2005} and Lempel-Ziv source coding~\cite{lonardi2007}. These source-channel coding schemes share the underlying idea of approximating the (optimum) maximum \textit{a posteriori} (MAP) decoder by using certain properties of the source statistics.

\begin{figure*}[t]
  \centering 
  \includegraphics[width=.8\textwidth]{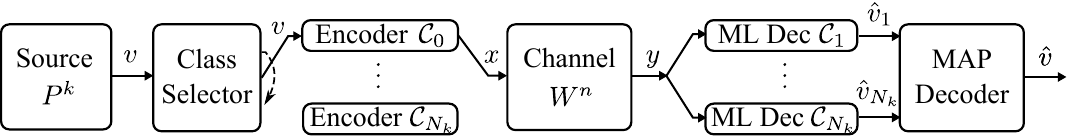}
  \caption{Block diagram of the multi-class source-channel coding scheme.  Class $\#0$ is reserved for a declared error, and it is not decoded at the receiver.\label{fig:multi-class}}
\end{figure*}

In this paper, we analyze an almost-lossless source-channel coding scheme in which source messages are assigned to disjoint classes and encoded by codes that depend on the class index. Under MAP decoding, this scheme attains the joint source-channel reliability function in the cases where it is known to be tight~\cite{tauste14}. We are interested in characterizing the performance of this coding scheme under simpler, sub-optimal decoding.
First, we process the channel output in parallel for each class using a bank of maximum likelihood (ML) decoders. Then, the decoded message is selected from the outputs of the ML decoders based on a MAP criterion. While this construction fails to achieve the best performance of joint source-channel coding, it presents a smaller complexity for a fixed number of classes. This scheme is shown to improve on the error exponent of separate coding, and, as the number of classes increases, to approach the error exponent of joint source-channel coding~\cite{Csis80}.

The proposed coding scheme can be interpreted as based on unequal error protection (UEP). The most probable messages are encoded with low-rate channel codes, and hence they receive an increased protection against channel errors. Analogously, less probable messages are assigned to classes that receive less protection against channel errors. UEP can also be implemented via an alternative coding scheme in which codewords are divided in two parts: a prefix that identifies which class the message belongs to, and a payload which encodes the message within the class. When the number of classes grows sub-exponentially with the block-length, the prefix encodes an information source of effective zero rate. Even in this case, for channels with no zero-error capacity, the prefix length is required to grow linearly with the code length. Therefore, this prefix scheme incurs a loss in exponent (see discussion after Lemma~2 in~\cite{Csis80}) and a loss in finite-length performance~\cite[Sec. II.C]{shkel15}. 

Shkel, Tan and Stark also proposed an alternative UEP coding scheme in~\cite{shkel15}. In their scheme, codewords belonging to different classes generally have different minimum distance, hence UEP is guaranteed via code construction. In contrast, in our scheme we do not require an UEP codebook. Codes for different classes are selected and optimized independently. Instead, UEP is achieved at the decoding stage by giving priority to some classes over the others. As a result, the source-channel code proposed  here can be designed and implemented with reduced complexity using existing source and channel codes, as shown with several examples.

The structure of the paper is as follows. In \refS{model} the system model and our multi-class source channel coding scheme are introduced. \refS{analysis} presents a random-coding analysis of this scheme. \refS{design} validates these results by means of simulation of a reduced complexity implementation based on LDPC codes, and \refS{conclusion} concludes the paper.

\section{System Model and Coding Scheme} \label{sec:model}

We consider the transmission of a length-$k$ discrete memoryless source
over a memoryless channel using length-$n$ block codes. We define $t\triangleq\frac{k}{n}$. The source output $\v=(v_1,\dots, v_k)\in \Vc^k$, where $\Vc$ is a discrete alphabet, is distributed according to $\Pv(\v)=\prod_{i=1}^{k}\pv(v_i)$, $\v=(v_1,\dots, v_k)\in \Vc^k$, where $\pv(v)$ is the source symbol distribution. Without loss of generality, we assume that $\pv(v)>0$ for all $\v$; if $\pv(v) = 0$ for some $v$, we define a new source without this symbol. The channel input $\x=(x_1,\dots, x_n)\in \Xc^n$ and output $\y=(y_1,\dots, y_n)\in \Yc^n$, where $\Xc$ and $\Yc$ respectively denote the input and output alphabet, are related via a channel law $\Pyx(\y|\x)=\prod_{i=1}^{n}\pyx(y_i|x_i)$, where $\pyx(y|x)$ denotes the channel transition probability. For the sake of clarity, in the following, we consider discrete channels. The analysis carries over to the case of continuous output alphabets, replacing the corresponding sums by integrals.

A source-channel code is defined by an encoder and a decoder. The encoder maps the message $\v$ to a length-$n$ codeword $\x(\v)$. Based on the channel output $\y$, the decoder selects a message $\hat\v(\y)$. When clear from context, we avoid writing the dependence of the decoder output on the channel output explicitly. Throughout the paper, random variables will be denoted by capital letters and the specific values they take on are
denoted by the corresponding lower case letters.
The error probability of a source-channel code is thus given by
\begin{align}
  \epsilon_n &= \Pr\bigl\{ \V \neq \hat\V \bigr\}.
\end{align}
We characterize this probability in terms of error exponents. An exponent $E(P,W,t) > 0$ is to said to be achievable if there exists a sequence of codes with $n=1,2,\ldots$, and $k=1,2,\ldots$, whose error probabilities $\epsilon_n$
satisfy
\begin{equation}\label{jscc_exponent_def}
\epsilon_n \leq e^{-n E(P,W,t)+o(n)}, 
\end{equation}
where $o(n)$ is a sequence such that $\lim_{n\to\infty} \frac{o(n)}{n}=0$.
The supremum of all achievable exponents $E(P,W,t)$ is usually referred to as reliability function.

Our coding scheme splits the source-message set in subsets, and use concatenated source and channel codes for each subset. At the receiver, each channel code is decoded in parallel, and the final output is selected based on the MAP criterion. A block diagram of this scheme is shown in \refF{multi-class}.

For each $k$, we define a partition $\Pc_k$ of the source-message set $\Vc^k$ into $N_k+1$ disjoint subsets $\Ac_{i}^{k}$, $i=0,1,\ldots,N_k$. We shall refer to these subsets as \emph{classes}. Sometimes, we consider sequences of sources, channels and partitions where $N_k$ grows with $k$. The asymptotic number of classes as $k\to\infty$ is $N \triangleq \lim_{k\to\infty} N_k$, hence $N \in \NN \cup \{\infty\}$.
More specifically, we consider partitions in which source messages are assigned to classes depending on their probability,
\begin{align} 
  \Ac_{i}^{k} = \left\{ \v \,\big|\, \gamma_{i}^k < \Pv(\v) \leq \gamma_{i+1}^k  \right\}, \
  i=0,\ldots,N_k,
   \label{eqn:Aidef}
\end{align}
with $0=\gamma_0^k \leq \gamma_1^k \leq \ldots \leq\gamma_{N_k+1}^k=1$.
Since the sets $\Ac_{i}^{k}$ are unions of type classes, $N_k$ grows (at most) subexponentially in $k$. We define the rate of each class as 
\begin{align}
R_i \triangleq \frac{1}{n} \log\bigl|\Ac_{i}^{k}\bigr|,\ i=0,\ldots,N_k.
   \label{eqn:Ridef}
\end{align}

All the messages in the class $\Ac_{0}^{k}$ are encoded with the same codeword $\x(\v) = \x_0$ and are assumed to lead to a decoding error. For each remaining class $\Ac_{i}^{k}$, messages are encoded with a channel code  $\Cc_i$ of rate $R_i$.
At the receiver, we use a two-step decoder (see \refF{multi-class}).
For each class $\Ac_{i}^{k}$, $i=1,\ldots,N_k$, the $i$-th ML decoder selects a message $\hat{\v}_i$ in $\Ac_{i}^{k}$ as
\begin{align}\label{eqn:MLdec}
\hat{\v}_i = \arg \max_{\v \in \Ac_{i}^{k}} \Pyx\bigl(\y | \x(\v)\bigr).
\end{align}
Next, the decoder selects from the set $\{\hat{\v}_i\}_{i=1}^{N_k}$, the source message with largest MAP decoding metric.
That is, the final output is $\hat{\v}=\hat{\v}_{\hat{\imath}}$, where the class index selected by the MAP decoder corresponds to
\begin{align}\label{eqn:MAPindex}
  \hat{\imath} &= \arg \max_{i=1,\ldots,N_k}                           
    q(\hat{\v}_i,\y),
\end{align}
where $q(\v,\y) \triangleq \Pv(\v) \Pyx\bigl(\y | \x(\v)\bigr)$.


\section{Error Exponent Analysis} \label{sec:analysis}


To analyze the random-coding error exponent of the scheme described in \refS{model}, we define three different error events. The first occurs when a source message belongs to the set $\Ac_{0}^{k}$ (source error); the second occurs when, for a source message belonging to class $\Ac_{i}^{k}$, the $i$-th ML decoder makes an error (ML error); and the third occurs when the $i$-th ML decoder output is correct but the MAP decoder makes an error (MAP error). More precisely, these three error events are defined respectively as
\begin{align}
\Ec_{\text{S}} &\triangleq \bigl\{\v\in\Ac_{0}^{k}\bigr\},\\
\Ec_{\text{ML}}(i) &\triangleq \bigl\{\v\in\Ac_{i}^{k} ,\, \hat{\v}_i \neq \v \bigr\},\\
\Ec_{\text{MAP}}(i) &\triangleq \bigl\{ \v\in\Ac_{i}^{k} ,\, \hat{\v}_i = \v ,\, \hat{\imath}\neq i \bigr\}.
\end{align}
Using that these error events are disjoint, we write the error probability as 
\begin{align} 
\epsilon_n
&= \Pr\left\{\Ec_{\text{S}} \cup \left(\bigcup_{i=1}^{N_k} \Ec_{\text{ML}}(i)\right) \cup \left(\bigcup_{i=1}^{N_k}\Ec_{\text{MAP}}(i)\right) \right\}
\label{eqn:errorprob0}\\
&= \Pr\bigl\{ \V \in \Ac_{0}^{k} \bigr\}
 + \sum_{i=1}^{N_k} \Pr\bigl\{\V\in\Ac_{i}^{k} ,\, \hat{\V}_i \neq \V \bigr\}
 +  \sum_{i=1}^{N_k} \Pr\bigl\{\V\in\Ac_{i}^{k},\, \hat{\V}_i=\V ,\, \hat{I}\neq i \bigr\}. \label{eqn:errorprob1}
\end{align}

To lower-bound  the error exponent, we start by upper-bounding every term in the third summand in~\eqref{eqn:errorprob1} as
\begin{align}
\Pr\bigl\{ \V\in\Ac_{i}^{k},\, \hat{\V}_i=\V,\, \hat{I}\neq i \bigr\}
&= \Pr\bigl\{ \hat{I} \neq i\,\big|\,\V\in\Ac_{i}^{k},\, \hat{\V}_i = \V \bigr\}\Pr\bigl\{\V\in\Ac_{i}^{k},\, \hat{\V}_i = \V \bigr\} \label{eqn:conderrorprob0}\\
&\leq \Pr\bigl\{ \hat{I} \neq i\,\big|\,\V\in\Ac_{i}^{k},\, \hat{\V}_i = \V \bigr\}\Pr\bigl\{\V\in\Ac_{i}^{k} \bigr\}, \label{eqn:conderrorprob1}
\end{align}
where \refE{conderrorprob0} follows from the chain rule, and \refE{conderrorprob1} by 
upper-bounding $\Pr\bigl\{\V\in\Ac_{i}^{k},\, \hat{\V}_i = \V \bigr\}$ by $\Pr\bigl\{\V\in\Ac_{i}^{k}\bigr\}$.
Using the MAP decoding rule in \refE{MAPindex}, the first factor in the right-hand side of \refE{conderrorprob1} can be upper-bounded as
\begin{align}
\Pr\bigl\{ \hat{I} \neq i\,\big|\,\V\in\Ac_{i}^{k},\, \hat{\V}_i = \V \bigr\}
&\leq \Pr\biggl\{q(\hat{\V}_i,\Y)  
    \leq \max_{j=1,\ldots,N_k, j\neq i} q(\hat{\V}_j,\Y)\;\Big|\;\V\in\Ac_{i}^{k} ,\, \hat{\V}_i = \V \biggr\} \label{eqn:conderrorprobMAPbound1a}\\
&\leq \Pr\biggl\{ q(\V,\Y)  
  \leq \max_{\bar\v\neq\V,\bar\v\notin\Ac_{0}^{k}} q(\bar\v,\Y) \,\Big|\,\V\in\Ac_{i}^{k} \biggr\},\label{eqn:conderrorprobMAPbound2}
\end{align}
where \refE{conderrorprobMAPbound1a} follows from \refE{MAPindex} by assuming that ties are decoded as errors, and \refE{conderrorprobMAPbound2} follows by applying the condition $\hat{\V}_i = \V$ and by enlarging the set of source messages over which the maximum is computed.

Substituting \refE{conderrorprob1} and \refE{conderrorprobMAPbound2} in \refE{errorprob1}, via the chain rule, yields
\begin{align} 
\epsilon_n
  &\leq \Pr\{ \V \in \Ac_{0}^{k} \}
+ \sum_{i=1}^{N_k} \Pr\bigl\{ \V\in\Ac_{i}^{k} ,\, \hat{\V}_i \neq \V \bigr\}
\notag\\&\;\;\;
+ \Pr\biggl\{ \V\notin\Ac_{0}^{k},\, q(\V,\Y)  
  \leq \max_{\bar\v\neq\V,\bar\v\notin\Ac_{0}^{k}} q(\bar\v,\Y)
  \biggr\}. \label{eqn:errorprob2}
\end{align}

In the following, we find useful to define the channel coding and source coding exponents. For $\rho \geq 0$ and $\px$ an arbitrary distribution over $\Xc$ let the Gallager's channel and source functions be given by
\begin{equation} \label{eqn:def_E0}
  \Eo(\rho, \px) \triangleq -\log \sum_{y} \left(\sum_{x}  \px(x) \pyx(y|x)^{\frac{1}{1+\rho}} \right)^{1+\rho},
\end{equation}
and
\begin{equation} \label{eqn:def_Es}
  \Es(\rho) \triangleq \log \left( \sum_{v} \pv(v)^{\frac{1}{1+\rho}} \right)^{1+\rho},
\end{equation}
respectively.
For channel coding alone, the random-coding exponent at rate $R$ for an input distribution $\px$ is achievable and it is given by~\cite{Gall68}
 \begin{align}
\Er(R, \px) = \max_{\rho\in[0,1]} \,\Bigl\{\Eo(\rho, \px)-\rho R \Bigr\}. \label{channel_exponent}
\end{align}
For source coding alone, the reliability function of a source $\pv$ at rate $R$,
denoted by $e(R)$, is given by~\cite{Jel68}
\begin{align}
e(R) = \sup_{\rho\geq 0} \bigl\{\rho R - \Es(\rho)\bigr\}.\label{source_exponent}
\end{align}

We upper-bound \refE{errorprob2} via a random-coding argument.
For every $k,n$, we assign a distribution $\px_i(x)$ to each class $\Ac_{i}^{k}$, $i=0,\ldots,N_k$, and randomly generate a codeword $\x(\v)$ according to $\px_i^n(\x) \triangleq \prod_{j=1}^{n}\px_i(x_j)$ for each source message $\v\in\Ac_{i}^{k}$ and each $i=1,\ldots,N_k$. For the class $\Ac_{0}^{k}$, we select a symbol distribution $\px_0$ that assigns mass $1$ to a predetermined null symbol. Then, its Gallager function satisfies that $\Eo(\rho_0,\px_0) = 0$ for any $\rho_0 \in [0,1]$. We also define $R_{N+1} \triangleq 0$ such that $e\left(\frac{R_{N+1}}{t}\right) = 0$. The next result follows from \refE{errorprob2} using the exponential bounds \cite[Th. 5.2]{Jel68}, \cite[Th. 5.6.1]{Gall68} and \cite[Th. 1]{tauste14}.

\begin{theorem}\label{thm:multi-class-exponent}
There exists a sequence of codes, partitions and decoders as defined in \refS{model} that achieves the exponent
\begin{align} 
  \min_{i=0,\ldots,N} \Biggl\{E_r(R_i,\px_i) + t e\biggl(\frac{R_{i+1}}{t}\biggr)\Biggr\}, \label{eqn:multi-class-exponent}
\end{align}
where $N = \lim_{k\to\infty} N_k$. Furthermore, for the set of rates $\{R_i\}$ maximizing \refE{multi-class-exponent} there exists a one-to-one relationship between each $R_i$ and the corresponding threshold $\gamma_{i}$ (see Lemma \ref{lem:paramEsi} in Appendix \ref{apx:multi-class-exponent}).
\end{theorem}
\begin{IEEEproof}
See Appendix \ref{apx:multi-class-exponent}. 
\end{IEEEproof} 


Under certain assumptions the lower bound in Theorem~\ref{thm:multi-class-exponent} coincides with an upper bound to the error exponent derived in \cite[Th.~2]{multijsccisit14} for the family of codes described in \refS{model}.
This is the case for a given class of channels (such as  the binary symmetric channel, binary erasure channel or phase-shift-keying modulated additive white Gaussian noise channel (AWGN)), when the intermediate rates optimizing \refE{multi-class-exponent} are above the critical rate of the channel and the codes $\Cc_1,\ldots,\Cc_{N_k}$ are linear.
While this converse result only applies to a class of codes and channels, it shows that in these cases there is no loss in exponent by considering the bound in Theorem~\ref{thm:multi-class-exponent}.

Further analysis involves optimization over rates $R_{i}$ (i.e., thresholds $\gamma_i$) and distributions $\px_i$, $i=1,\ldots,N$. The bound in Theorem~\ref{thm:multi-class-exponent} can be relaxed to obtain an alternative expression. We define $\Eo(\rho) \triangleq \max_{\px} \Eo(\rho, \px)$.

\begin{theorem}\label{thm:multi-class-exponent-bis}
There exists a sequence of codes, partitions and decoders defined in \refS{model} with $N \geq 2$ that achieves the exponent
\begin{align}
\max_{R' \geq R \geq 0}\min\Biggl\{\;& 
     \max_{\rho\geq0} \bigl\{ \rho R' - t\Es(\rho) \bigr\},
     \notag\\ &
     \max_{\bar\rho\in[0,1]} \left\{\Eo(\bar\rho)
              - t \Es(\bar\rho)
              - \bar\rho\frac{R'  - R}{N-1} \right\},
     \notag\\ &
     \max_{\bar{\bar\rho}\in[0,1]} \bigl\{
     \Eo(\bar{\bar\rho}) - \bar{\bar\rho} R \bigr\}
     \Biggr\},\label{eqn:multi-class-exponent-bis}
\end{align}
Moreover, the rate of the $i$-th class in the partition is
\begin{equation} \label{eqn:optRi}
 R_i = R + (i-1) \frac{R' - R}{N-1}, \qquad i=1,\ldots,N,
\end{equation}
where $R$ and $R'$ are the values optimizing \refE{multi-class-exponent-bis}.
\end{theorem}
\begin{IEEEproof}
See Appendix \ref{apx:multi-class-exponent-bis}.
\end{IEEEproof} 

The bound in Theorem \ref{thm:multi-class-exponent-bis} is simple to evaluate since it only involves the well known functions $\Es(\cdot)$ and $\Eo(\cdot)$, and the optimization is performed over a fixed number of parameters ($\rho,\bar\rho,\bar{\bar\rho}$, $R$ and $R'$), independent of $N$.
Furthermore, as we verify next with an example, it is sometimes indistinguishable from the bound in Theorem~\ref{thm:multi-class-exponent}.

For $N=1$, we have that $\Er(R_0,\px_0)=0$ and $e(R_2)=0$. Optimizing \refE{multi-class-exponent} over intermediate rate $R = R_1$ and distribution $\px_1$, Theorem \ref{thm:multi-class-exponent} recovers the separate source-channel exponent~\cite{Csis80},
\begin{align}
\max_{R \geq 0}\min\biggl\{
      E_r(R) , \; te\biggl(\frac{R}{t}\biggr)
     \biggr\},\label{eqn:separate-exponent}
\end{align}
where $\Er(R) \triangleq \max_{\px} \Er(R, \px)$.

Let $N_{k}$ grow (subexponentially) with $k$ in such a way that $\lim_{k\to\infty} N_k = \infty$. For this discussion only, we allow $R$ and $R'$ to depend on $k$ as $R_k$ and $R_k'$, respectively. Let us choose the sequences $R_k$ and $R_k'$ such that $\lim_{k\to\infty} R_k = 0$, $\lim_{k\to\infty} R_k'=\infty$ and $\lim_{k\to\infty} \frac{R_k' - R_k}{N_k-1} = 0$, i.e., $R_k'=o(N_k)$.
In this case, the first and last terms within the minimization in \refE{multi-class-exponent-bis} become irrelevant and the bound in Theorem \ref{thm:multi-class-exponent-bis} recovers Gallager's source-channel error exponent~\cite[p. 534, Prob.~5.16]{Gall68},
\begin{align}
     \max_{\rho\in[0,1]}  \bigl\{\Eo(\rho)
              - t \Es(\rho) \bigr\}.
     \label{eqn:joint-exponent}
\end{align}

In several cases of interest, the exponent \refE{joint-exponent} coincides with the joint source-channel reliability function. However, for specific source and channel pairs the following exponent gives a tighter bound to the reliability function~\cite{Csis80,Zhong06},
\begin{equation}
  \min_{R\geq0} \left\{E_r(R) + t e\left(\frac{R}{t}\right)\right\}
    = \max_{\rho\in[0,1]} \bigl\{\barEo(\rho) - t \Es(\rho)\bigr\},
  \label{eqn:joint-concave-hull-exponent}
\end{equation}
where $\barEo(\rho)$ denotes the concave hull of $\Eo(\rho)$, defined pointwise as the supremum  over convex combinations of any two values of the function $\Eo(\rho)$~\cite[p. 36]{convex-rockafellar}. 
While the bound in Theorem \ref{thm:multi-class-exponent-bis} does not attain \refE{joint-concave-hull-exponent}, this error exponent can be recovered from Theorem \ref{thm:multi-class-exponent} by identifying the classes with the source-type classes $\Psf_{i}$, $i=1,\ldots,N_k$. In this case, $R_i = t H(\Psf_{i})$ and $R_{i+1} = t H(\Psf_{i+1})$ become infinitely close to each other and they uniformly cover the interval $\bigl[0,t\log(|\Vc|)\bigr]$ for $i=1,2,\ldots$. As a result, \refE{multi-class-exponent} recovers the left-hand side of \refE{joint-concave-hull-exponent}. This shows that the gap between the bounds in Theorems \ref{thm:multi-class-exponent} and \ref{thm:multi-class-exponent-bis} can be strictly positive.

\subsection{Example}

\begin{figure}[t]
  \centering  \includegraphics[width=.6\columnwidth]{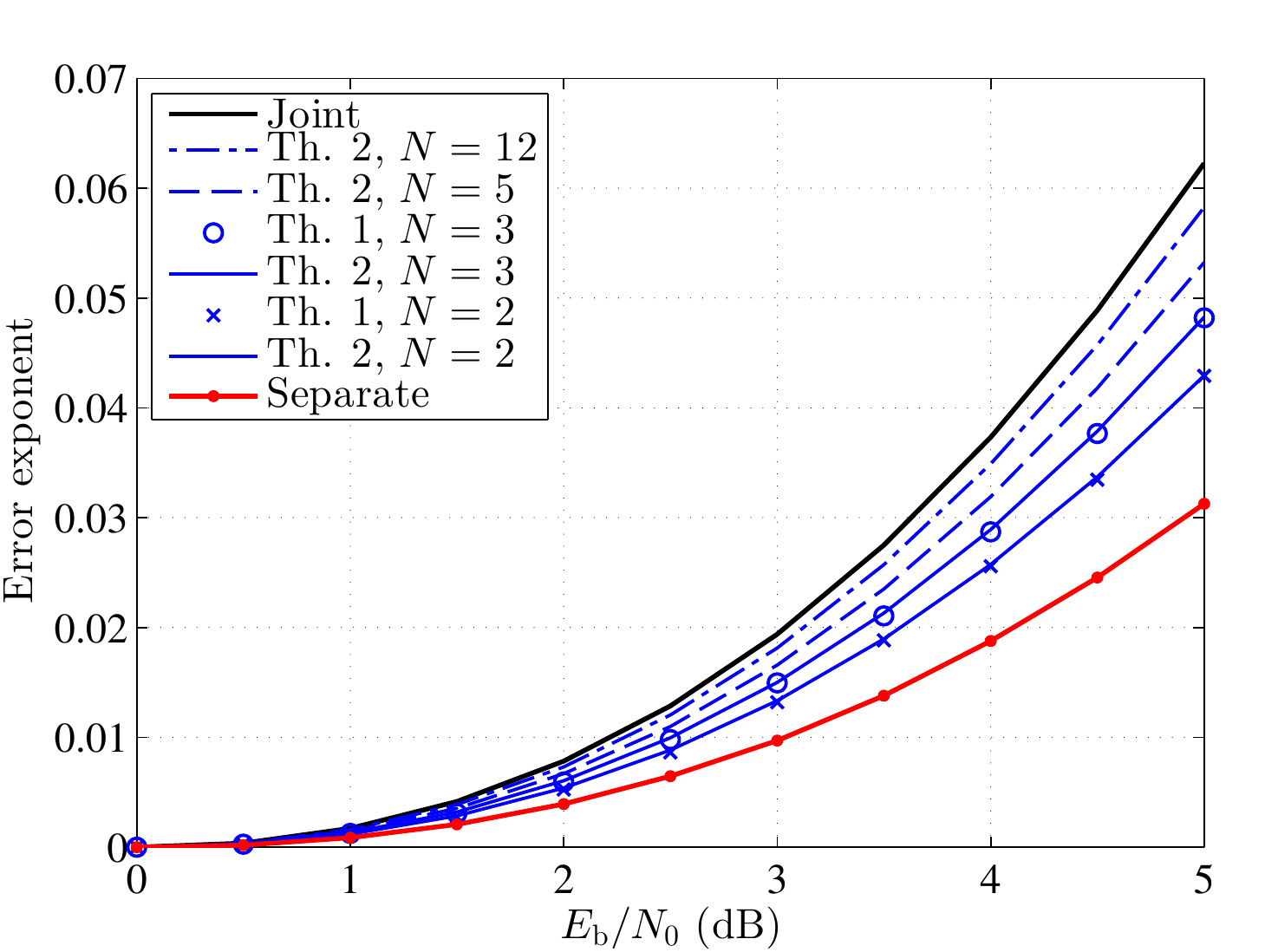}
\caption{ Error exponent bounds. BMS with $\pv(1)=0.1$ transmitted over a binary input AWGN channel for $t=1$.\label{fig:exponents}}
\end{figure}

A binary memoryless source (BMS) with parameter $p \triangleq \pv(1) \leq 1/2$ is to be transmitted over a binary-input AWGN channel with signal-to-noise ratio (SNR) ${E_{\rm s}}/{N_0}$.
For comparison purposes, we normalize ${E_{\rm s}}/{N_0}$ with respect to the number of transmitted {\em information} bits if the source were compressed to entropy, i.e.~$tH(V)$. Let $h_2(p) = -p\log_{2}p-(1-p)\log_{2}(1-p)$ denote the binary entropy function in bits.  We define a signal-to-noise ratio per source bit ${E_{\rm b}}/{N_0}$ as
\begin{equation} \label{SNRbit}
  \frac{E_{\rm b}}{N_{0}} \triangleq \frac{n}{k h_2(p)} \frac{E_{\rm s}}{N_{0}}.
\end{equation}
\refFig{exponents} shows the achievable error exponents for different coding schemes as a function of ${E_{\rm b}}/{N_0}$ in decibels.
The error exponents in the figure correspond to separate source-channel coding \refE{separate-exponent}, joint source-channel coding \refE{joint-exponent}, and the multi-class scheme with $N=2,3,5,12$.
The bound in Theorem \ref{thm:multi-class-exponent} has been 
optimized over the parameters $\rho_i$, $i=0,\ldots,N$, and thresholds $\gamma_i$, $i=1,\ldots,N$.
The bound in Theorem \ref{thm:multi-class-exponent-bis} has been optimized over the parameters $\rho,\bar\rho,\bar{\bar\rho}$, $R$ and $R'$. In both cases the channel input distribution has been chosen to be equiprobable.
From the figure we can see that the bound in  Theorem \ref{thm:multi-class-exponent} and the relaxed version in Theorem \ref{thm:multi-class-exponent-bis} coincide for $N=2,3$. For $N=2$, the multi-class scheme shows a $0.4$-$0.7$ dB improvement over separate source coding, with just a small increase in complexity.
Moreover, from the curves for $N=2,3,5,12$ we can see that the multi-class construction approaches the joint source-channel error exponent as the number of classes increases, confirming the results of Theorems~\ref{thm:multi-class-exponent} and~\ref{thm:multi-class-exponent-bis} (since \refE{joint-exponent} and \refE{joint-concave-hull-exponent} coincide for this example).

\section{Practical Code Design}  \label{sec:design}

Based on the proposed scheme, we now design a practical joint source-channel code for the transmission of a BMS with $\pv(1) \le 1/2$ over a binary-input AWGN channel. In particular, we consider a two-class code composed of a fixed-to-variable lossless source code followed by two linear codes with different rates. The lossless source code corresponds to the class selector in \refF{multi-class}. The ML decoders in \refF{multi-class} can be replaced by using standard quasi-ML decoders. This fixed-to-variable-to-fixed source-channel code allows a simple implementation using existing source and channel codes:

First, the length-$k$ binary source sequence is encoded using a fixed-to-variable coding scheme that assigns shorter codewords to the most probable messages, i.e., messages with smallest Hamming weight. Two examples are enumerative~\cite{Cover1973} and arithmetic coding~\cite{WNC1987}. For a source message $\v$, the length of the source codeword $L(\v)$ determines which code will be used to encode each source message. Since the source code is assumed lossless, this is equivalent to assigning source messages to classes based on their probability.

As channel codes we consider two linear $(n,k_i)$-codes $\Cc_i$, $i=1,2$.
If $L(\v) \le k_1$ the channel code $\mathcal C_1$ is used for transmission, otherwise, if $L(\v) \le k_2$ the second code, $\mathcal C_2$, is used. If $L(\v) > k_2$ an arbitrary codeword is used and a source coding error is reported. In this coding scheme, $k_i-L(\v)$ leftover bits may appear due to a mismatch between the source and channel code, $i=1,2$.
These bits can be used to include additional redundancy checks (see~\cite{jsccturbo14} for details), however, we set them to zero for the sake of simplicity.
Due to these leftover bits, we do not use all the codewords belonging to each of the channel codes, in contrast to the analysis in \refS{analysis}.
However, in general, $k_i \approx \log_2\bigl|\Ac_{i}^{k}\bigr|$, $i=1,2$, and the performance loss is small.


At the decoder, two ML (or quasi-ML) parallel decoding attempts are performed, one of each channel code. Both decoder outputs are then checked to verify whether they are valid source sequences. If only one of the two outputs is a valid source message, the corresponding data are used. If both decoders fail, a predetermined message, for example the all-zero data sequence, is used. Finally, if both source decoders report success, the message with larger a posteriori likelihood is selected.

\subsection{Code Optimization}

The specific pair of $(n,k_i)$-codes depends on the signal-to-noise ratio ${E_{\rm b}}/{N_0}$. Obviously, the choice of the code rates and of the codes themselves is critical for the system performance. If the block length $n$ is small, we can obtain a set of good channel codes with different coding rates using techniques from, e.g.~\cite{HBJK2009,BK1997,isit2013}. Then, for each ${E_{\rm b}}/{N_0}$ the best pair of codes from this set can be selected by simulating the system performance. While this optimization procedure is feasible for short block lengths, it becomes computationally intractable as the block length or rate granularity grow large.

In these cases, we may resort to the error exponents deriven in \refS{analysis} to estimate the optimal coding rate pair. To this end we compute the optimal rates $R_1$ and $R_2$ from either Theorem \ref{thm:multi-class-exponent} or Theorem \ref{thm:multi-class-exponent-bis}, and select two codes of rates  $R_1$ and $R_2$.
Since the exponential behavior dominates for large block lengths, these rates become asymptotically optimal as the block length grows large.
As we will see in the simulations section, Theorems \ref{thm:multi-class-exponent} and \ref{thm:multi-class-exponent-bis} give a good approximation of the optimal coding rates for moderate block lengths ($n \approx 1000$).

\subsection{Lower bound on the error probability}

We derive a lower bound on the error probability of a two-class linear coding scheme for a BMS. This lower bound will serve as a benchmark to the performance of practical codes.

Disregarding the last summand in \refE{errorprob1} we lower bound the error probability of a given code as
\begin{align}
\epsilon_n
  \geq &\Pr\bigl\{ \V \in \Ac_{0}^{k} \bigr\}
+ \sum_{i=1,2} \Pr\bigl\{ \V\in\Ac_{i}^{k},\, \hat{\V}_i \neq \V \bigr\}
\label{eqn:errorprob3}\\
  = &\Pr\bigl\{ \V \in \Ac_{0}^{k} \bigr\}
+ \sum_{i=1,2} \Pr\bigl\{ \V\in\Ac_{i}^{k} \bigr\} \Pr\bigl\{ \hat{\V}_i \neq \V \,\big|\, \V\in\Ac_{i}^{k} \bigr\}.
\label{eqn:errorprob4}
\end{align}

A lower bound on the error probability of a channel code of rate $R$ is given by Shannon's sphere-packing bound~\cite{Shannon1959}. 

Let codewords be distributed over the surface of an $n$-dimensional hypersphere with squared radius $E = n E_{\rm s}$ and centered at the origin of coordinates. Let $\theta$ be the half-angle of a cone with vertex at the origin and with axis going through one arbitrary codeword. We let $Q(\theta)$ denote the probability that such codeword be moved outside the cone by effect of the Gaussian noise. We choose $\theta_{n,R}$ such that the solid angle subtended by a cone of half-angle $\theta_{n,R}$ is equal to $\Omega_n/2^{nR}$, where $\Omega_n$ is the surface of the $n$-dimensional hypersphere. Then, $Q(\theta_{n,R_i})$ is a lower bound on the error probability of the $i$-th (length-$n$) linear codes under ML decoding (when ties are resolved randomly), i.e.,
\begin{align} \label{eqn:shannon_linearbound}
\Pr\bigl\{ \hat{\V}_i \neq \V \,\big|\, \V\in\Ac_{i}^{k} \bigr\} \geq Q(\theta_{n,R_i}),\  \; i=1,2.
\end{align}
This bound is accurate for low SNRs and relatively short codes~\cite{igal2006}. In order to compute \refE{shannon_linearbound} we shall use the approximation from \cite{VB1999}, known to be accurate for error probabilities below $0.1$.

For a BMS with $p = \pv(1) \le 1/2$, it is possible to obtain a closed-form expression for the source terms $\Pr\bigl\{ \V\in\Ac_{i}^{k} \bigr\}$, $i=0,1,2$. Consider a class $\Ac_{i}^{k}$ composed by the sequences with Hamming weights $w \in [w_1, w_2]$,  where $w_1$ and $w_2$ are two arbitrary integers. Then, it follows that
\begin{align}
\Pr\bigl\{ \V\in\Ac_{i}^{k} \bigr\} &= B_{k,p}(w_1,w_2),\label{eqn:BMSprob}
\end{align}
where we defined
\begin{align}
B_{k,p}(w_1,w_2)&\triangleq\sum_{w=w_1}^{w_2}\binom{k}{w} p^w (1-p)^{k-w}.
\end{align}

The best coding strategy is to encode the sequences of Hamming weight $w\in\{0,\ldots,w_1\}$ with the first (lower-rate) channel code and the sequences of weight $w\in\{w_1+1,\ldots, w_2\}$ with the second (higher-rate) code.
All other sequences are transmitted by some fixed codeword which leads to decoding error.
Therefore, using \refE{shannon_linearbound} and \refE{BMSprob} in \refE{errorprob4}, we obtain the following result.

\begin{theorem}Consider a length-$k$ BMS with $p = \pv(1) \le 1/2$ to be transmitted over a binary-input AWGN channel using a length-$n$ block code. The error probability of any two-class scheme using linear channel codes and ML decoding (with randomly resolved ties), is lower bounded as
\begin{align} \label{eqn:shannon_lowerbound}
\epsilon_n
  \geq \min_{\substack{w_1=0,\dots,k,\\w_2=w_1+1,\dots,k}} \Bigl\{ &B_{k,p}(0,w_1) Q\bigl(\theta_{n,R(0,w_1)}\bigr)
   \nonumber\\&
   + B_{k,p}(w_1+1,w_2) 
  Q\bigl(\theta_{n,R(w_1+1,w_2)}\bigr)
   \nonumber\\&
+ B_{k,p}(w_2+1,k) \Bigr\},
\end{align}
where the rate $R(w_1,w_2)$ is given by
\begin{equation}
R(w_1,w_2)=\frac{1}{n} \left\lceil  \log_{2} \sum_{w=w_1}^{w_2} \binom{k}{w} \right\rceil.
\end{equation}

\end{theorem}

\subsection{Simulation Results}

In this subsection we show simulation results for different implementations of a two-class scheme in short and moderate block length scenarios. The source probability is fixed to $\pv(1)=0.1$.

\begin{figure}[t]
  \centering
  \includegraphics[width=.6\columnwidth]{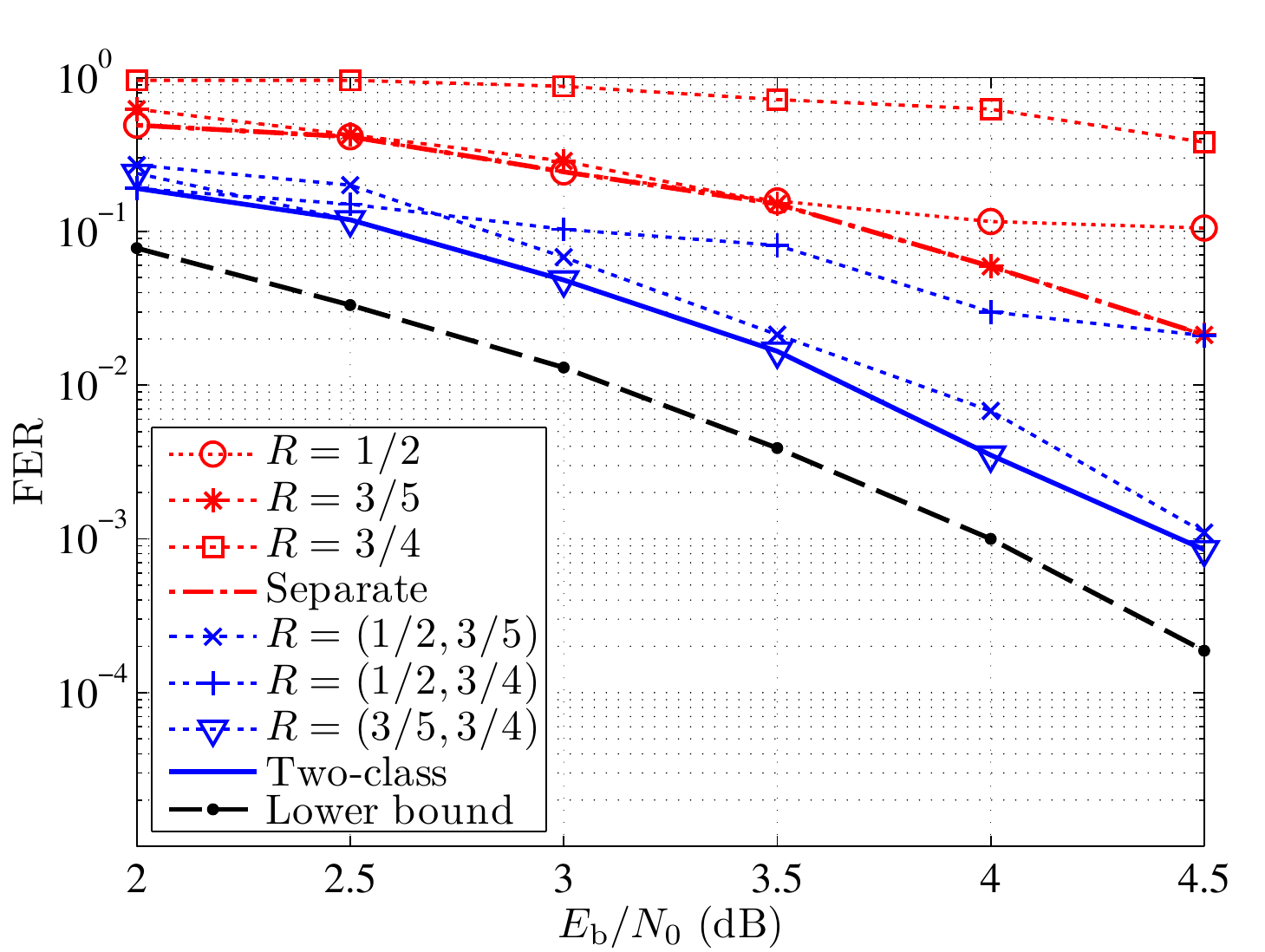}
 \caption{Enumerative + TB coding, $n=100$, $k=80$. Frame error rate for separate  and two-class source-channel coding.\label{fig:JSCC100}}
\end{figure}
\begin{table}[t]
  \centering
\caption{Enumerative + TB coding, $n=100$, $k=80$. Optimal rate pairs $(R_1, R_2)$ for a two-class coding scheme.\label{tab:JSCC100}}
\begin{tabular}{c c c}
${E_{\rm b}}/{N_0}$& Simulation & Asymptotic analysis\\\hline
$2$ dB & $(0.5, 0.75)$ & $(0.447, 0.475)$\\
$3$ dB & $(0.6, 0.75)$ & $(0.481, 0.522)$\\
$4$ dB & $(0.6, 0.75)$ & $(0.516, 0.569)$\\
\end{tabular}
\end{table}

\subsubsection{Short block length scenario ($k=80$, $n\approx 100$)}\refFig{JSCC100} shows the simulated frame error rate (FER) performance of an implementation using tail-biting codes and ML decoding. As source code we use an enumerative coding scheme and as channel codes we have chosen a family of tail-biting (TB) codes of rates $R= 1/2, 3/5$ and $3/4$. The code of rate $R=1/2$ was taken from \cite{BJKS2002}, and the codes of rates $3/5$ and $3/4$ where chosen by doing a short search for high-rate convolutional codes using techniques from \cite{HBJK2009,BK1997}. Among the most efficient ML decoding algorithms we have selected BEAST \cite{BEAST2004dec} which allows ML decoding for codes of length $100$ with acceptable complexity.
The curves ``Separate'' and ``Two-class'' show the best performance obtained within the corresponding family of codes. The two-class scheme outperforms separate coding by about $1$ dB, in agreement with the values predicted by the random coding analysis.
Also, from the figure we see that the lower bound \refE{shannon_lowerbound} can be used to predict not only the gain value but also the best error probability.

Table \ref{tab:JSCC100} shows the best code rate pairs obtained for different values of ${E_{\rm b}}/{N_0}$ in this scenario. The table compares the values obtained by simulating pairs of TB codes $R= 1/2, 3/5$ and $3/4$ with the asymptotic results obtained from \refE{optRi} in Theorem \ref{thm:multi-class-exponent-bis}.
We can see that there is a discrepancy  between simulation and asymptotic analysis, due to the short block length considered or possibly to the coarse granularity of the coding rates. 

\begin{figure}[t]
  \centering
  \includegraphics[width=.6\columnwidth]{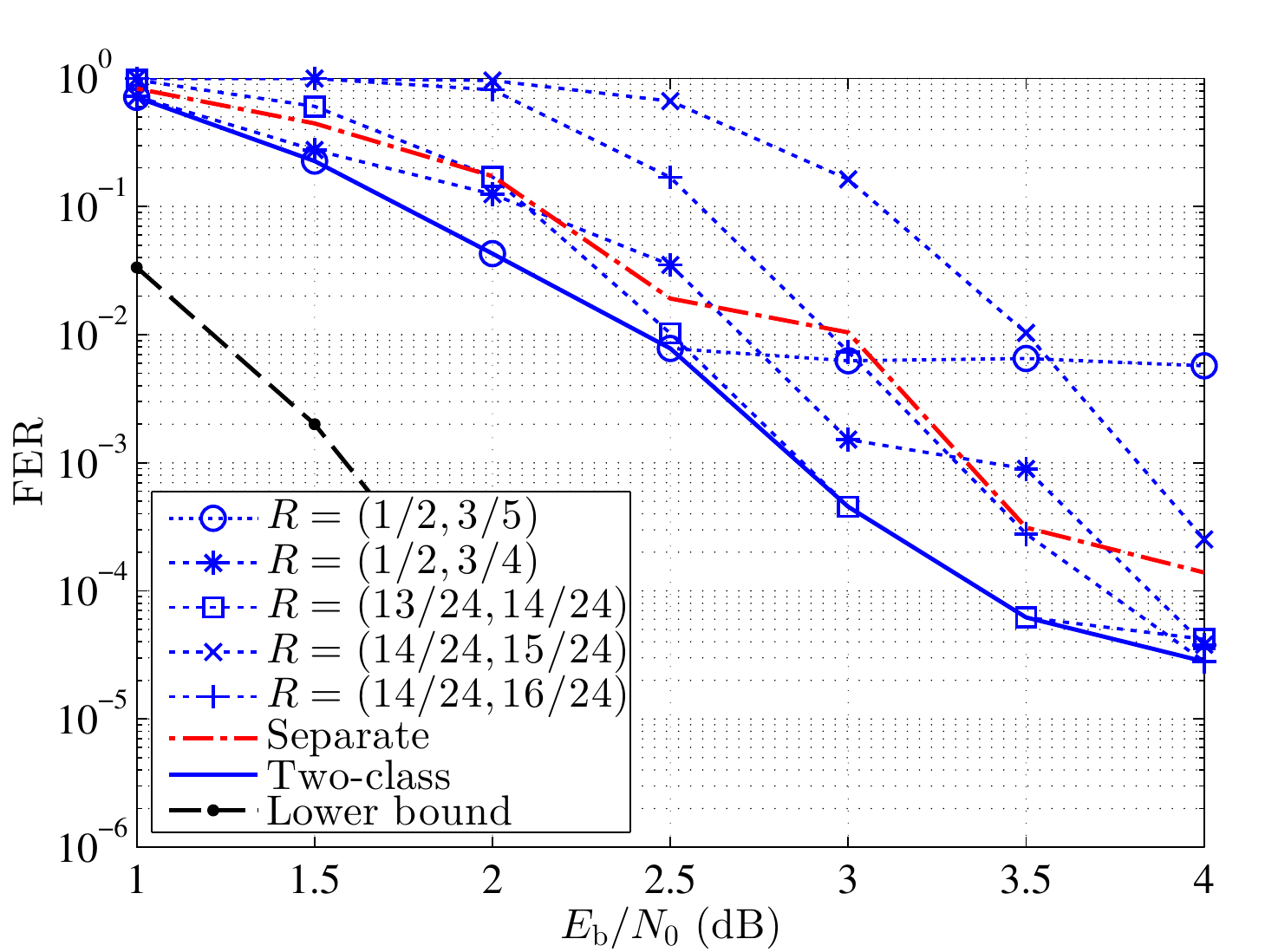}
 \caption{Enumerative + LDPC coding, $n=1008$, $k=1000$. Frame error rate for separate and two-class source-channel coding.\label{fig:JSCC1000}}
\end{figure}
\begin{table}[t]
  \centering
\caption{Enumerative + LDPC coding, $n=1008$, $k=1000$. Optimal rate pairs $(R_1, R_2)$ for a two-class coding scheme.\label{tab:JSCC1000}}
\begin{tabular}{c c c}
${E_{\rm b}}/{N_0}$& Simulation & Asymptotic analysis\\\hline
$1$ dB & $(0.5, 0.6)$ & $(0.499, 0.511)$\\
$2$ dB & $(0.5, 0.6)$ & $(0.536, 0.561)$\\
$3$ dB & $(0.542, 0.583)$ & $(0.575, 0.612)$\\
$4$ dB & $(0.583, 0.667)$ & $(0.614, 0.664)$\\
\end{tabular}
\end{table}

\subsubsection{Moderate block length scenario ($k=1000$, $n\approx 1000$)}\refFig{JSCC1000} shows the FER for an implementation using LDPC codes and iterative decoding.
We use enumerative source coding and a family of quasi-cyclic (QC) LDPC codes as channel codes. In particular we consider a set of codes with $24$-column base matrix and coding rates $R=12/24,13/24,\ldots,16/24$.
For constructing these parity-check matrices we used the optimization algorithm from \cite{isit2013}. The only exception is the code of rate $R=18/24$ which is borrowed from \cite[code~A]{STD802}. The decoding algorithm is stopped after $50$ iterations of belief propagation decoding and we require at least 50 block error events for each simulated point.

Each separate source-channel code presents an error floor due to the effect of the source coding error events which do not depend on the channel SNR. This phenomenon results in a staggered behavior both of separate and the two-class codes curves in \refF{JSCC1000}. For clarity, no individual separate source-channel curves have been plotted, but only the best performance within the family (``Separate''). We can see that the two-class scheme (``Two-class''), optimized for each SNR point, outperforms separate coding by $0.4$-$0.7$ dB. In this case the gap to the lower bound \refE{shannon_lowerbound} is larger with respect to that in \refF{JSCC100} because of the suboptimal decoding algorithm, with performance far from ML decoding.

Table \ref{tab:JSCC1000} shows the best code rate pairs in this scenario.
We observe a better agreement between asymptotic results and simulation results. This is due to the larger block length, that makes the asymptotic approximations more accurate. This fact justifies the use of the asymptotic analysis from \refS{analysis} to guide the design of good finite-length codes.

\section{Concluding Remarks} \label{sec:conclusion}
In this paper we have presented a source-channel coding scheme in which the source messages are divided into classes based on their probability and a channel code and ML decoding is used for each of the classes.
We have shown that the overall scheme outperforms separate source-channel coding and approaches the performance of joint source-channel coding as the number of classes increases.

The multi-class scheme can be implemented using existing source and channel codes with reduced complexity. Simulation results for a binary memoryless source transmitted over a binary input additive Gaussian channel show that using two classes offers a 0.5-1.0 dB gain compared to separate source-channel coding. This is consistent with the theoretically predicted values. Moreover, analytical results have been shown to offer a practical guideline to the design of finite-length source-channel codes in the memoryless setting. While the analysis is restricted to memoryless sources and channels, the multi-class scheme could be easily implemented for sources and channels with memory by using appropriate source and channel codes.


\appendices 

\section{Proof of Theorem \ref{thm:multi-class-exponent}}
\label{apx:multi-class-exponent}

In order to prove Theorem \ref{thm:multi-class-exponent} we start by introducing a number of properties of the partition of the source message set. The main proof is then included in Section \ref{apx:multi-class-exponent-proof} of this appendix.

\subsection{Properties of the partition $\bigl\{\Ac_{i}^{k}\bigr\}$ in \refE{Aidef}}

Let us define the function
\begin{align} 
  \Essub{i}(\rho) &\triangleq \lim_{k\to\infty} \frac{1}{k} 
\log \left( \sum_{\v \in \Ac_{i}^{k}} \Pv(\v)^{\frac{1}{1+\rho}} \right)^{{1+\rho}},
  \label{eqn:defEsi}
\end{align}
which takes over the role of Gallager's source function $\Es(\cdot)$ when dealing with multiple classes (see, e.g.,~\cite{tauste14}). 
In principle, the functions $\Essub{i}(\cdot)$ are difficult to evaluate, since they involve summing over an exponential number of terms (one for each sequence) and the computation of a limit.
The following result provides a simple characterization of $\Essub{i}(\cdot)$ for a sequence of partitions of the form \refE{Aidef}. We denote the derivative of $\Es(\rho)$ evaluated at $\rho$ as
\begin{align}
\Es'(\rho) \triangleq
  \left.\frac{\partial \Es(\bar\rho)}{\partial \bar\rho}\right|_{\bar\rho=\rho}
\end{align}
and we define the tilted distribution
\begin{align}
\pv_{\sigma}(v) \triangleq \frac{\pv(v)^{\sigma}}
                  {\sum_{\bar v} \pv(\bar v)^{\sigma}}. \label{eqn:pv_sigma}
\end{align}

\begin{lemma}\label{lem:Esi} 
Consider a sequence of memoryless sources $\pv^k$ and partitions $\bigl\{\Ac_{i}^{k}\bigr\}$ in \refE{Aidef}, $k=1,2,\ldots$. Then, for any $\rho\in\RR$, $\gamma_i \leq \max_{v} \pv(v)$ and $\gamma_{i+1} > \min_v \pv(v)$,
\begin{align} \label{eqn:Esi}
&\Essub{i}(\rho) = \begin{cases}
         \Es(\rho_{i}^{\star}) + (\rho-\rho_{i}^{\star}) \Es'(\rho_{i}^{\star}),\qquad\qquad\;\; \tfrac{1}{1+\rho}<\tfrac{1}{1+\rho_{i}^{\star}},\\
         \Es(\rho),\qquad\qquad\qquad\qquad\  \tfrac{1}{1+\rho_{i}^{\star}} \leq \tfrac{1}{1+\rho} \leq \tfrac{1}{1+\rho_{i+1}^{\star}},\\
         \Es(\rho_{i+1}^{\star}) + (\rho-\rho_{i+1}^{\star}) \Es'(\rho_{i+1}^{\star}),\,\quad \tfrac{1}{1+\rho} > \tfrac{1}{1+\rho_{i+1}^{\star}},
      \end{cases}
\end{align}
where $\rho_{i}^{\star}$,  $i=0,\ldots,N+1$, are given by the solution to the implicit equation $\sum_{v} \pv_{\frac{1}{1+\rho_i^{\star}}}(v) \log \pv(v) = \log\gamma_i$
as long as $\min_{v} \pv(v) \leq \gamma_i \leq \max_{v} \pv(v)$. When $\gamma_i < \min_{v} \pv(v)$, $\rho_i^{\star} = -1_{-}$ and for $\gamma_i > \max_{v} \pv(v)$, $\rho_i^{\star} = -1_{+}$.

For $\gamma_i > \max_{v} \pv(v)$ or $\gamma_{i+1} \leq \min_v \pv(v)$,
the $i$-th class is empty and $\Essub{i}(\rho) = -\infty$.
\end{lemma}
\begin{IEEEproof}
See Appendix \ref{apx:Esi}.
\end{IEEEproof}

In principle, the values of $\rho_{i}^{\star}$ appearing in Lemma \ref{lem:Esi} can be negative. If we restrict ourselves to the range $\rho \geq 0$, the thresholds yielding negative values of $\rho_{i}^{\star}$ are uninteresting to us, since they correspond to  classes that never dominate the exponent. Therefore, for the present work, we may restrict the value of the thresholds $\gamma_i$ to satisfy $\sum_{v} \frac{1}{|\Vc|} \log \pv(v) \leq \log \gamma_i \leq \sum_{v} \pv(v) \log \pv(v)$, $i=1,\ldots,N$. In this case, the three regions in $\rho$ appearing in \refE{Esi} can be equivalently written as $\bigl\{\rho>\rho_{i}^{\star}\bigr\}$, $\bigl\{\rho_{i+1}^{\star} \leq \rho \leq \rho_{i}^{\star}\bigr\}$, and $\bigl\{\rho<\rho_{i+1}^{\star}\bigr\}$, respectively, with $\rho_0^{\star}=\infty$ and $\rho_{N+1}^{\star}=0$.

\begin{figure}[t]
  \begin{center}
  \includegraphics[width=.45\columnwidth]{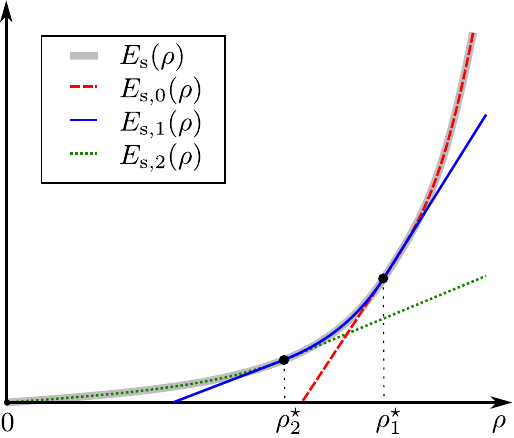}
	\caption{Example of the characterization in Lemma \ref{lem:Esi}  of the $\Essub{i}(\cdot)$ functions with three classes ($N=2$).}
	\label{fig:Esi}
  \end{center}
\end{figure}

An example of the characterization in Lemma \ref{lem:Esi} for $\rho\geq0$ is shown in \refF{Esi} for a three-class partition.
We observe that $\Essub{i}(\rho)$ is equal to $\Es(\rho)$ for the interval $\rho_{i+1}^{\star} \leq \rho \leq \rho_{i}^{\star}$, and corresponds to a straight line tangent to $\Es(\rho)$ out of those intervals. Since the thresholds $\gamma_0$ and $\gamma_{N+1}$ are fixed to $0$ and $1$, respectively, then $\rho_0^{\star}=\infty$ and $\rho_{N+1}^{\star}=0$.
For the remaining thresholds, we can obtain any finite value of $\rho_i^{\star}\in[0,\infty)$ by appropriately choosing the threshold $\gamma_i$, $i=1,\ldots,N$, between 
$\exp \bigl(\sum_{v} \frac{1}{|\Vc|} \log \pv(v) \bigr)$ and $\exp \bigl(\sum_{v} \pv(v) \log \pv(v) \bigr)$.

\begin{lemma}\label{lem:paramEsi} 
For a sequence of memoryless sources $\pv^k$ and partitions $\bigl\{\Ac_{i}^{k}\bigr\}$ in \refE{Aidef}, $k=1,2,\ldots$, each threshold $\min_v \pv(v) \leq \gamma_i \leq \max_{v} \pv(v)$ in \refE{Aidef} univocally determines the corresponding coding rate $R_i$ in \refE{Ridef} for each $i=1,\ldots,N$, with $N=\lim_{k\to\infty} N_k$. 
In particular,
  \begin{align}\label{eqn:paramEsi}
   R_i = t \Es'(\rho_i^{\star})
\end{align}
where $\rho_{i}^{\star}$ is given by the solution to the implicit equation
$\sum_{v} \pv_{\rho_i^{\star}}(v) \log \pv(v) = \log\gamma_i$.
\end{lemma}
\begin{IEEEproof}
See Appendix \ref{apx:paramEsi}.
\end{IEEEproof}

Lemmas \ref{lem:Esi} and \ref{lem:paramEsi} imply that, asymptotically, it is equivalent to optimize the partition over either the set of thresholds $\{\gamma_{i}\}$ or over the rates $\{R_{i}\}$. Furthermore, they provide an alternative representation of the asymptotic probability of the set $\Ac_{i}^{k}$, as shown by the next result.

\begin{lemma}\label{lem:limPrAki} 
Consider a sequence of memoryless sources $\pv^k$ and partitions $\bigl\{\Ac_{i}^{k}\bigr\}$ in \refE{Aidef}, $k=1,2,\ldots$. When $\log \gamma_{i+1} \leq \sum_{v} \pv(v) \log \pv(v)$, $i=1,\ldots,N$, it holds that
\begin{align} 
  \lim_{k\to\infty} \frac{1}{k} \log \Biggl( \sum_{\v \in \Ac_{i}^{k}} \Pv(\v) \Biggr)
     &= -e\biggl(\frac{R_{i+1}}{t}\biggr). \label{eqn:limPrAki}
\end{align}
\end{lemma}
\begin{IEEEproof}
From~\refE{defEsi} we have that
\begin{align} 
  \lim_{k\to\infty} \frac{1}{k} \log \Biggl( \sum_{\v \in \Ac_{i}^{k}} \Pv(\v)  \Biggr)
   &= \Essub{i}(0)  \label{eqn:limPrAki-0} \\
  &= \Es(\rho_{i+1}^{\star}) -\rho_{i+1}^{\star} \Es'(\rho_{i+1}^{\star}) \label{eqn:limPrAki-1} \\
  &= \max_{\rho \geq 0}\,\biggl\{\Es(\rho) -\rho \frac{R_{i+1}}{t}\biggr\}, \label{eqn:limPrAki-2}
\end{align}
where \refE{limPrAki-1} follows from \refE{Esi} given the assumptions in the lemma implying $\rho_{i+1}^{\star} \geq 0$, and in~\refE{limPrAki-2} we used Lemma~\ref{lem:paramEsi} and the fact that $\rho_{i+1}^{\star}$ is the point where $\Es(\rho)$ has slope $\frac{R_{i+1}}{t}$, i.e., it maximizes the quantity in brackets. The result thus follows from \refE{limPrAki-2} by using the definition~\eqref{source_exponent} of the error exponent of a discrete memoryless source compressed to rate $\frac{R_{i+1}}{t}$.
\end{IEEEproof}

Then, the asymptotic coding rate of the $i$-th class is uniquely determined by the lower threshold $\gamma_{i}$ defining this class, as shown in Lemma~\ref{lem:paramEsi}.
Similarly, combining Lemma~\ref{lem:paramEsi} and Lemma~\ref{lem:limPrAki}, we obtain that the exponent of the probability of the $i$-th class is determined by the upper threshold $\gamma_{i+1}$.

\subsection{Proof of Theorem \ref{thm:multi-class-exponent}} 
\label{apx:multi-class-exponent-proof}

We now proceed with the proof of Theorem \ref{thm:multi-class-exponent}.
Under the assumption that the number of classes $N_k$ behaves sub-exponentially in $k$, the error exponent is given by the minimum of the individual exponents of each of the summands in \refE{errorprob2}, namely
\begin{align} 
- \lim_{n\to\infty} \frac{1}{n} \log \epsilon_n
=  \min\biggl\{&- \lim_{n\to\infty} \frac{1}{n} \log \Pr\{ \V \in \Ac_{0}^{k} \},
\notag\\&
\min_{i=1,\dotsc,N} - \lim_{n\to\infty} \frac{1}{n} \log \Pr\bigl\{ \V\in\Ac_{i}^{k} ,\, \hat{\V}_i \neq \V \bigr\}, \notag \\ &
-\!\lim_{n\to\infty} \frac{1}{n}\!\log \Pr\Bigl\{ \V\notin\Ac_{0}^{k},\, q(\V,\Y)  
  \leq\!\max_{\bar\v\neq\V,\bar\v\notin\Ac_{0}^{k}}\!q(\bar\v,\Y)
  \Bigr\}\!\biggr\}\!. \label{eq:error_exp1}
\end{align}
We next analyze each of the terms in the minimum separately.

As we discussed after Lemma~\ref{lem:Esi}, we consider partitions with thresholds $\gamma_i$ satisfying $\sum_{v} \frac{1}{|\Vc|} \log \pv(v) \leq \log \gamma_i \leq \sum_{v} \pv(v) \log \pv(v)$, $i=1,\ldots,N$.  Then, Lemma~\ref{lem:limPrAki} yields the exponent of the first term in the minimum in~\eqref{eq:error_exp1}, that is
\begin{align}
  - \lim_{n\to\infty} \frac{1}{n} \log \Pr\{ \V \in \Ac_{0}^{k} \} 
   &= te\biggl(\frac{R_{1}}{t}\biggr)\label{eq:logPrA0}.
\end{align}

We now upper bound the second term in~\eqref{eq:error_exp1}. First, we use the chain rule to express the probability, for $i=1,\ldots,N$, as
\begin{align} 
  \Pr\bigl\{  \V\in\Ac_{i}^{k} ,\, \hat{\V}_i \neq \V \bigr\}
  &= \Pr\bigl\{ \hat{\V}_i \neq \V | \V\in\Ac_{i}^{k} \bigr\} \Pr\bigl\{ \V\in\Ac_{i}^{k} \bigr\}.\label{eq:logPrJSCi}
\end{align}
The first factor corresponds to the error probability of a channel coding problem with $M_i$ messages transmitted over a channel $\Pyx$. We can lower-bound its exponent in terms of the random-coding exponent for input distribution $Q_i$. For each each class $\Ac_{i}^{k}$, $i=1,\ldots,N$, there exists a code $\Cc_i$ whose error probability over the memoryless channel $\pyx$ satisfies~\mbox{\cite[Th. 5.6.1]{Gall68}}
\begin{align} 
-\lim_{n\to\infty} \frac{1}{n} \Pr\bigl\{ \hat{\V}_i \neq \V | \V\in\Ac_{i}^{k} \bigr\}
   &\geq \max_{\rho_i\in[0,1]}\bigl\{\Eo(\rho_i, \pyx, \px_i) - \rho_i R_i\bigr\},\label{eqn:boundCh6} \\
   &= E_r(R_i, \px_i). \label{eqn:boundCh8}
\end{align}
As in~\eqref{eq:logPrA0}, the exponent of the second factor in~\eqref{eq:logPrJSCi} is
\begin{align}
  - \lim_{n\to\infty} \frac{1}{n} \log \Pr\{ \V \in \Ac_{i}^{k} \} &= te\biggl(\frac{R_{i+1}}{t}\biggr)\label{eq:logPrAi}.
\end{align} 
Combining~\eqref{eqn:boundCh8} and~\eqref{eq:logPrAi} we thus obtain
\begin{align} 
  - \lim_{n\to\infty} \frac{1}{n} \log \Pr\bigl\{ \V\in\Ac_{i}^{k} ,\, \hat{\V}_i \neq \V \bigr\}    &\geq E_r(R_i,Q_i) + te\biggl(\frac{R_{i+1}}{t}\biggr), \quad i=1,\ldots,N.\label{eqn:boundCh7}
\end{align}

Finally, we identify the last term in~\refE{errorprob2} as the error exponent of a specific joint source-channel coding problem, where the source message probabilities do not add up to 1. In the random-coding argument, codewords are generated according to a class-dependent input distribution $Q_i$, $i=1,\ldots,N$. We can thus use~\cite[Th. 1]{tauste14} to bound the exponent
\begin{align}
-\!\lim_{n\to\infty}\frac{1}{n} \log&\Pr\biggl\{ q(\V,\Y)  
  \leq\!\max_{\bar\v\neq\V,\bar\v\notin\Ac_{0}^{k}}\!q(\bar\v,\Y) ,\, \V\notin\Ac_{0}^{k}
  \biggr\}\geq \min_{i=1,\dotsc,N} \Bigl\{
               \Eo\bigl(\bar\rho_i,\pyx,\px_{i}\bigr) 
               - t\Essub{i}(\bar\rho_i)\Bigr\},
      \label{eqn:boundJ1}
\end{align}
for any $\bar\rho_{i}\in[0,1]$. Here we used that the proof of \cite[Th. 1]{tauste14} is valid also for defective source message probabilities.

From Lemma~\ref{lem:Esi}, we infer that the source function $\Essub{i}(\rho)$ is non-decreasing, convex and with a non-decreasing derivative. Moreover, Lemma~\ref{lem:paramEsi} shows that the derivative approaches the limiting value $\frac{R_i}{t}$ as $\rho\to\infty$. Therefore, the source function $\Essub{i}(\rho)$ satisfies the following simple upper bound for non-negative $\rho$
\begin{align}
  \Essub{i}(\rho) &\leq \Essub{i}(0) + \rho \frac{R_i}{t}  \\
  &= -e\biggl(\frac{R_{i+1}}{t}\biggr) + \rho \frac{R_i}{t},
    \label{eqn:Esilowerbound-1}
\end{align}
where we used~\refE{limPrAki-1}. Substituting~\eqref{eqn:Esilowerbound-1} in the right-hand side~\eqref{eqn:boundJ1} we obtain 
\begin{align}
   &\Eo\bigl(\bar\rho_i,\pyx,\px_{i}\bigr)
               - t\Essub{i}(\bar\rho_i)
               \geq  \Eo\bigl(\bar\rho_i,\pyx,\px_{i}\bigr) - \bar\rho_i R_i + te\biggl(\frac{R_{i+1}}{t}\biggr).
\end{align}
Since this inequality holds for arbitrary $\bar\rho_{i}\in[0,1]$ and input distribution $\px_i$, we conclude that for each value of $i = 1,\dotsc,N$, the corresponding exponent in~\eqref{eqn:boundJ1} is lower-bounded by  the exponent in~\eqref{eqn:boundCh7}. Hence, this term can be omitted in the minimum in~\eqref{eq:error_exp1}. 

Finally, we observe that $\px_0$ satisfies $E_r(R,\px_0) = 0$ for any rate $R$. Then, from~\eqref{eq:error_exp1}, using the intermediate results \eqref{eq:logPrA0}, with   $te\bigr(\frac{R_{1}}{t} \bigr)$ replaced by $te\bigr(\frac{R_{1}}{t} \bigr) + E_r(R_1,\px_0)$, and~\refE{boundCh7} we get the desired
\begin{align} 
- \lim_{n\to\infty} \frac{1}{n} \log \epsilon_n
  \geq  \min_{i=0,\dotsc,N}  \Biggr\{ E_r(R_i,\px_i) + te\biggr(\frac{R_{i+1}}{t} \biggr)   \Biggr\}. \label{eqn:expbound-1}
\end{align}

Lemma~\ref{lem:paramEsi} shows that for $t\Es'(0) \leq R_i \leq  \lim_{\rho\to\infty} t\Es'(\rho)$ the correspondence between $\gamma_{i}$ and $R_i$ is one-to-one. Since the set $\{R_i\}$ that maximizes the right-hand side of \refE{expbound-1} is always in this range, we conclude that it is  asymptotically equivalent to optimize the partition over thresholds $\{\gamma_i\}$ or rates $\{R_i\}$.

\subsection{Proof of Lemma \ref{lem:Esi}}
\label{apx:Esi}

For $\sigma\in\RR$ and $k=1,2,\ldots$, let us define the random variable $Z_{\sigma,k} \triangleq \log \Pv(\V)$ with underlying distribution
\begin{align}
\Pv_{\sigma}(\v) \triangleq \frac{\Pv(\v)^{\sigma}}
                  {\sum_{\bar\v} \Pv(\bar\v)^{\sigma}}.
\end{align}
This distribution is the multi-letter version of~\refE{pv_sigma}. The asymptotic normalized log-moment generating function of $Z_{\sigma,k}$ is given by
\begin{align}
\kappa_\sigma(\tau) 
  &\triangleq \lim_{k\to\infty} \frac{1}{k} \log \Ex\bigl[ e^{\tau Z_{\sigma,k}} \bigr]\\
  &=
  \log \left( \frac {\sum_{v} \pv(v)^{\sigma+\tau}}
  {\sum_{\bar v} \pv(\bar v)^{\sigma} } \right).\label{eqn:kappataudef}
\end{align}

It follows that
\begin{align}
\Lambda_i(\sigma) 
&\triangleq\lim_{k\to\infty}\frac{1}{k}\log\Biggl( \sum_{\v \in \Ac_{i}^{k}} \Pv(\v)^{\sigma} \Biggr)\\
&= \lim_{k\to\infty}\frac{1}{k}\log\Biggl( \sum_{\bar\v} \Pv(\bar\v)^{\sigma}\Biggr)
 + \lim_{k\to\infty}\frac{1}{k}\log\Biggl( \sum_{\v \in \Ac_{i}^{k}} \Pv_{\sigma}(\v) \Biggr)\\
&= \log\left( \sum\nolimits_{v} \pv(v)^{\sigma} \right)
+\!\lim_{k\to\infty}\frac{1}{k}\log\Bigl( \Pr\bigl\{\log\gamma_i^k<Z_{\sigma,k}\leq\log\gamma_{i+1}^k\bigr\} \Bigr).
\end{align}

Applying the Gartner-Ellis theorem~\cite[Th. II.6.1]{ellis-1985} to the term $\Pr\bigl\{\log\gamma_i^k<Z_{\sigma,k}\leq\log\gamma_{i+1}^k\bigr\}$, and given the smoothness properties of $\kappa_\sigma(\tau)$ in \refE{kappataudef}, we obtain
\begin{align}
    \Lambda_i(\sigma) = \sup_{\log\gamma_i \leq r \leq \log\gamma_{i+1}} \inf_{\tau} \; \Phi(r, \tau),
      \label{eqn:lem1-opt}
\end{align}
where
\begin{align}
\Phi(r, \tau) 
  &\triangleq \log\left( \sum\nolimits_{v} \pv(v)^{\sigma} \right) 
              - \bigl( r\tau - \kappa_\sigma(\tau) \bigr)\label{eqn:lem1-phi-def-1}\\
  &= - r\tau + \log \left( \sum\nolimits_{v} \pv(v)^{\sigma+\tau} \right). \label{eqn:lem1-phi-def-2}
\end{align}

The function $\Phi(r, \tau)$ is differentiable in $\CC^2$ and that its Hessian is given by
\begin{align}
  \nabla^2_\Phi(r, \tau) = \left[ \begin{array}{cc}
                               0 & -1 \\
                               -1 & \frac{\partial^2 \Phi(r, \tau)}{(\partial \tau)^2}  \end{array} \right].
\end{align}
Hence, its determinant is $\bigl|\nabla^2_\Phi(r, \tau)\bigr| =  - 1 < 0$ and the solution of \refE{lem1-opt} is a saddle point provided that the constraints are non-active. By taking the derivative of $\Phi(r, \tau)$ with respect to $\tau$ and equating it to zero we obtain that for the optimal point it holds that
\begin{align}
  r = \sum\nolimits_{v} \pv_{\sigma+\tau}(v) \log \pv(v).
\label{eqn:lem1-condr}
\end{align}
By taking the derivative of $\Phi(r, \tau)$ with respect to $r$ and equating it to zero it follows that for the optimal point 
\begin{align}
  \tau = 0, \label{eqn:lem1-condtau}
\end{align}
provided that the constraints in \refE{lem1-opt} are non-active.

We translate the constraints on $r$ to the domain of $\sigma$.
Let $\sigma_{i}^{\star}$ be given by the solution to the implicit
equation
\begin{align}
  \sum\nolimits_{v} \pv_{\sigma_i^{\star}}(v) \log \pv(v) = \log\gamma_i,
  \label{eqn:defsigmaistar}
\end{align}
as long as  $\min_v \pv(v) \geq \gamma_i \geq \max_{v} \pv(v)$.
In case that $\gamma_i < \min_v \pv(v)$ then $\sigma_i^{\star}=-\infty$; if $\gamma_i > \max_{v} \pv(v)$, then $\sigma_i^{\star}=\infty$.
Using \refE{lem1-condr} and \refE{lem1-condtau}, the constraints in \refE{lem1-opt}, $\log\gamma_i \leq r \leq \log\gamma_{i+1}$, can be equivalently written as $\sigma_{i}^{\star} \leq \sigma \leq \sigma_{i+1}^{\star}$, $i=0,\ldots,N$.
\begin{enumerate}
\item
When $\sigma_{i}^{\star} \leq \sigma \leq \sigma_{i+1}^{\star}$ the constraints are non-active and the saddlepoint occurs at
\begin{align}
   r  = \sum\nolimits_{v} \pv_{\sigma}(v) \log \pv(v), \qquad \tau = 0.
\end{align}
Substituting these values in \refE{lem1-opt} we obtain
\begin{align}
    \Lambda_i(\sigma)
    &= \log \left( \sum\nolimits_{v} \pv(v)^{\sigma} \right).\label{eqn:lem1-case1}
\end{align}

\item
For $\sigma < \sigma_{i}^{\star}$, the optimal $r$ is given by
\begin{align}
r  = \log \gamma_{i} = \sum\nolimits_{v} \pv_{\sigma_{i}^{\star}}(v) \log \pv(v),
\end{align}
and using \refE{lem1-condr}, we obtain $\tau = \sigma_{i}^{\star} - \sigma$.
Substituting these values in \refE{lem1-opt} yields
\begin{align}
\Lambda_i(\sigma)
&= \left(\sigma -\sigma_{i}^{\star}\right) \sum\nolimits_{v} \pv_{\sigma_{i}^{\star}}(v) \log \pv(v) + \log \left( \sum\nolimits_{v} \pv(v)^{\sigma_{i}^{\star}} \right).
\label{eqn:lem1-case2}
\end{align}

\item
Proceeding in an analogous way to the previous case, for $\sigma > \sigma_{i+1}^{\star}$, we obtain
\begin{align}
\Lambda_i(\sigma)
&= \left(\sigma -\sigma_{i+1}^{\star}\right) \sum\nolimits_{v} \pv_{\sigma_{i+1}^{\star}}(v) \log \pv(v) + \log \left( \sum\nolimits_{v} \pv(v)^{\sigma_{i+1}^{\star}} \right).
\label{eqn:lem1-case3}
\end{align}
\end{enumerate}
Substituting \refE{lem1-case1}, \refE{lem1-case2} and \refE{lem1-case3}, $i=0,\ldots,N$, in the corresponding range of the parameter~$\sigma$, rearranging terms, we obtain
\begin{align} \label{eqn:Lambdai}
    \frac{1}{\sigma}\Lambda_i(\sigma) &= 
      \begin{cases}
      G(\sigma, \sigma_{i}^{\star}), &\sigma<\sigma_{i}^{\star},\\
         \tfrac{1}{\sigma} \log \Bigl( \sum\nolimits_{v} \pv(v)^{\sigma}\Bigr), &\sigma_{i}^{\star} \leq \sigma \leq \sigma_{i+1}^{\star},\\
      G(\sigma, \sigma_{i+1}^{\star}), &\sigma>\sigma_{i+1}^{\star},\\
      \end{cases}
\end{align}
where
\begin{align}
G(\sigma, s) &\triangleq \tfrac{1}{s} \log \Bigl( \sum\nolimits_{v} \pv(v)^{s} \Bigr)
 -\left(\tfrac{1}{\sigma} - \tfrac{1}{s}\right) \sum\nolimits_{v} \pv_{s}(v) \log \pv_{s}(v).
\end{align}
The expression $\tfrac{1}{\sigma}\Lambda_i(\sigma)$ in \refE{Lambdai} corresponds precisely with $\Essub{i}(\rho)$ when $\sigma = \tfrac{1}{1+\rho}$. Then, the result follows from the definition of $\Es(\rho)$ in \refE{def_Es}, using that
\begin{align}
 \Es'(\rho) = - \sum_{v} \pv_{\frac{1}{1+\rho}}(v) \log \pv_{\frac{1}{1+\rho}}(v).
\end{align}

\subsection{Proof of Lemma \ref{lem:paramEsi}}
\label{apx:paramEsi}

Using the characterization in Lemma \ref{lem:Esi} it follows that
\begin{align}
  \lim_{\rho\to\infty} \frac{1}{\rho} \Essub{i}(\rho)
  &=\lim_{\rho\to\infty} \frac{1}{\rho} \left(
           \Es(\rho_i^{\star}) + (\rho-\rho_i^{\star}) \Es'(\rho_i^{\star}) \right)
  \label{eqn:limEsi-1a}\\
  &= \Es'(\rho_i^{\star}),
  \label{eqn:limEsi-1b}
\end{align}
as long as $\rho_i^{\star}<\infty$. 

Also, using the definition \refE{defEsi} we have that
\begin{align}
  \lim_{\rho\to\infty}\frac{1}{\rho} \Essub{i}(\rho)
    &= \lim_{\rho\to\infty} \lim_{k\to\infty}
    \frac{1}{\rho k} 
    \log \left( \sum_{\v \in \Ac_{i}^{k}} \Pv(\v)^{\frac{1}{1+\rho}} \right)^{1+\rho}
\label{eqn:limEsi-2a}\\
    &= \lim_{k\to\infty} \lim_{\rho\to\infty}
    \frac{1}{\rho k} 
    \log \left( \sum_{\v \in \Ac_{i}^{k}} \Pv(\v)^{\frac{1}{1+\rho}} \right)^{1+\rho} \label{eqn:limEsi-2b}\\
    &= \lim_{k\to\infty}    \frac{1}{k} 
 \lim_{\rho\to\infty}
    \frac{1+\rho}{\rho} 
    \log \left( \sum_{\v \in \Ac_{i}^{k}} \Pv(\v)^{\frac{1}{1+\rho}} \right)
\label{eqn:limEsi-2c}\\
    &= \lim_{k\to\infty}    \frac{1}{k} 
    \log\,\bigl|\Ac_{i}^{k}\bigr|
\label{eqn:limEsi-2d}\\
    &= \frac{R_i}{t},
\label{eqn:limEsi-2e}
\end{align}
where in \refE{limEsi-2b} we applied the Moore-Osgood theorem~\cite[p. 619]{osgood29} since the expression
\begin{align}
  \frac{1}{\rho k} 
  \log \left( \sum_{\v \in \Ac_{i}^{k}} \Pv(\v)^{\frac{1}{1+\rho}} \right)^{1+\rho}
\label{eqn:limEsi-arg}
\end{align}
presents uniform convergence for each $k$ as
$\rho\to\infty$, and pointwise convergence as $k\to\infty$, as we show next. Then, using \refE{limEsi-1a}-\refE{limEsi-1b} and \refE{limEsi-2a}-\refE{limEsi-2e}, we obtain \refE{paramEsi}. The result thus follows from the definition of $\rho_{i}^{\star}$ in Lemma~\ref{lem:Esi}.

We show the convergence properties of \refE{limEsi-arg}. We write
\begin{align}
\frac{1}{k}\log \left( \sum_{\v \in \Ac_{i}^{k}} \Pv(\v)^{\frac{1}{1+\rho}} \right)^{{1+\rho}}\!-
    \frac{1}{k}\log\,\bigl|\Ac_{i}^{k}\bigr|
     &\leq   
     \frac{1}{k} \left(
    \log \left( \sum_{\v \in \Ac_{i}^{k}} 1^{\frac{1}{1+\rho}} \right)^{\frac{1+\rho}{\rho}} -
    \log\,\bigl|\Ac_{i}^{k}\bigr|\right)\\
    &=
     \frac{1}{k} \left(
    \log \, \bigl|\Ac_{i}^{k}\bigr|^{\frac{1+\rho}{\rho}}  -
    \log\,\bigl|\Ac_{i}^{k}\bigr| \right)\\
    &= \frac{1}{k\rho} 
     \log \,\bigl|\Ac_{i}^{k}\bigr| \\
    &= \frac{R_i}{t\rho}.
    \label{eqn:proof-MooreOsgood-1}
\end{align}
Similarly,
\begin{align}
\frac{1}{k} \log\,\bigl|\Ac_{i}^{k}\bigr|-
        \frac{1}{k} \log \left(\sum_{\v \in \Ac_{i}^{k}} \Pv(\v)^{\frac{1}{1+\rho}} \right)^{{1+\rho}}
&\leq 
     \frac{1}{k} \left(
     \log\,\bigl|\Ac_{i}^{k}\bigr|
     -
     \log \left( \sum_{\v \in \Ac_{i}^{k}} \left(\min_v \pv(v)\right)^{\frac{k}{1+\rho}} \right)^{\frac{1+\rho}{\rho}} \right)\\
    &=
     \frac{1}{k} \left(
     \log\,\bigl|\Ac_{i}^{k}\bigr|
    - \log\,\bigl|\Ac_{i}^{k}\bigr|^{\frac{1+\rho}{\rho}} 
    - \log \left( \min_v \pv(v)\right)^{\frac{k}{\rho}}
     \right)\\
    &= - \frac{1}{k\rho} 
    \log \, \bigl|\Ac_{i}^{k}\bigr| 
     - \frac{1}{\rho}  \log \, \min_v \pv(v)\\
     &= \frac{1}{\rho} \left(- \log \,\min_v \pv(v) - \frac{R_i}{t} \right).
    \label{eqn:proof-MooreOsgood-2}
\end{align}
Since \refE{proof-MooreOsgood-1} and \refE{proof-MooreOsgood-2} do not depend on $k$, \refE{limEsi-arg} presents uniform convergence with respect to $k$ as $\rho\to\infty$. Pointwise convergence of \refE{limEsi-arg} as $k\to\infty$ follows from \refE{Esi}.

\section{Proof of Theorem \ref{thm:multi-class-exponent-bis}}
\label{apx:multi-class-exponent-bis}

We start by writing~\eqref{eqn:multi-class-exponent} in dual form, that is, as explicit maximizations over parameters $\rho_i$ and $\bar\rho_i$, 
\begin{align} 
  - \lim_{n\to\infty} \frac{1}{n} \log \epsilon_n
  &\geq  \min_{i=0,\ldots,N} \Biggl\{\max_{\bar\rho_i\in[0,1]} \bigl\{E_0(\bar\rho_i,\px_i) -\bar\rho_i R_i\bigr\} + \max_{\rho_i\in[0,\infty)}  \bigl\{\rho_iR_{i+1} -tE_s(\rho_i)\bigr\}\Biggr\}. \label{eqn:expbound-2}
\end{align}
For $i = 0$ we have $E_r(R,\px_0) = 0$ and for $i = N$ we have $e\bigl(\frac{R_{N+1}}{t}\bigr) = 0$. In the range $i = 1,\dotsc,N-1$ we may fix $\rho_i = \bar\rho_i$ without violating the inequality in \refE{expbound-2}. Then, optimizing over $\px_i$, $i=1,\ldots,N$, we obtain
\begin{align} 
- \lim_{n\to\infty} \frac{1}{n} \log \epsilon_n
\geq 
\max_{R_1 \geq \ldots \geq R_{N} \geq 0}
\min\biggl\{\;&
     \max_{\rho_0\in[0,\infty)} \bigl\{ \rho_0 R_1 - t\Es(\rho_0) \bigr\},
     \notag\\ &
       \min_{i=1,\ldots,N-1} \max_{\bar\rho_i\in[0,1]} \bigl\{\Eo(\bar\rho_i)
              - t \Es(\bar\rho_{i})
              - \bar\rho_i (R_i  - R_{i+1}) \bigr\},
     \notag\\ &
     \max_{\bar\rho_{N}\in[0,1]} \bigl\{
     \Eo(\bar\rho_{N}) - \bar\rho_{N} R_{N} \bigr\}
     \biggr\}.\label{eqn:expbound-3}
\end{align}
Noting that the inner minimization in \refE{expbound-3} is maximized with respect to $\{R_i\}$ when $R_i  - R_{i+1}$ is constant, $i=1,\ldots,N-1$, the result follows.

\bibliographystyle{IEEEtran}
\bibliography{bib/references,bib/JSCC}

\end{document}